\documentclass[reprint,superscriptaddress, amsmath,amssymb, aps, pra, ]{revtex4-2}

\usepackage[usenames, dvipsnames]{color}
\usepackage{comment}
\usepackage{hyperref}
\usepackage{float}
\hypersetup{
    colorlinks=true,
    linkcolor=Blue,
    citecolor=Blue,
    filecolor=Blue,      
    urlcolor=Blue,
    pdftitle={Spinor droplets},
    }
\usepackage{graphicx}
\usepackage{physics}

\newcommand{\tildeOmega}{\tilde{\Omega}}

\def\U0{\tilde{U}(0)}

\def\a-kd{a^{\dagger}_{-k}}

\def\b-kd{b^{\dagger}_{-k}}

\def\sumk0{\sum_{\vec{k} \neq 0}}

\begin{document}

\title{Vortex Lattices in Strongly Confined Quantum Droplets}

\author{T.A. Yo\u{g}urt}
 \email{ayogurt@metu.edu.tr}
 \affiliation{%
Department of Physics, Middle East Technical University, Ankara, 06800, Turkey\\
}%

\author{U. Tanyeri}
 \affiliation{%
Department of Physics, Middle East Technical University, Ankara, 06800, Turkey\\
}%

\author{A. Kele\c{s}}
 \affiliation{%
Department of Physics, Middle East Technical University, Ankara, 06800, Turkey\\
}%

\author{M.\"O. Oktel}
\affiliation{Department of Physics, Bilkent University, Ankara, 06800, Turkey}

\date{\today}

\begin{abstract}
Bose mixture quantum droplets display a fascinating stability that relies on quantum fluctuations to prevent collapse driven by mean-field effects. Most droplet research focuses on untrapped or weakly trapped scenarios, where the droplets exhibit a liquid-like flat density profile. When weakly trapped droplets rotate, they usually respond through center-of-mass motion or splitting instability. Here, we study rapidly rotating droplets in the strong external confinement limit where the external potential prevents splitting and the center-of-mass excitation. We find that quantum droplets form a triangular vortex lattice as in single-component repulsive Bose-Einstein condensates (BEC), but the overall density follows the analytical Thomas-Fermi profile obtained from a cubic equation. We observe three significant differences between rapidly rotating droplets and repulsive BECs. First, the vortex core size changes markedly at finite density, visible in numerically obtained density profiles. 
We analytically estimate the vortex core sizes from the droplets' coherence length and find good agreement with the numerical results.
Second, the change in the density profile gives a slight but observable distortion to the lattice, which agrees with the distortion expected due to nonuniform superfluid density. Lastly, unlike a repulsive BEC, which expands substantially as the rotation frequency approaches the trapping frequency, rapidly rotating droplets show only a fractional change in their size. We argue that this last point can be used to create clouds with lower filling factors, which may facilitate reaching the elusive strongly correlated regime. 
\end{abstract}

\maketitle

\section{\label{sec: Introduction} INTRODUCTION}
Rotating a Bose-Einstein condensate (BEC) results in non-classical phenomena due to the irrotational nature of superfluid flow in the absence of density singularities \cite{pethick2008bose, pitaevskii2016bose}. A phase-coherent BEC carries angular momentum through quantized vortices, forming a regular vortex lattice to mimic solid body rotation. Single-component BECs with repulsive interactions were experimentally observed to create a  triangular lattice of singly quantized vortices when rotated in an isotropic harmonic trap \cite{2000_PRL_Dalibard_Vortex, 1999_PRL_Cornell_Vortex,2001_Ketterle_Science}. The rotation frequency and superfluid density dependence of the vortex core size, vortex density, and condensate radius of the rotating repulsive BECs are theoretically studied for both the Thomas-Fermi (TF) and the lowest Landau level (LLL) regimes \cite{2009_RMP_Fetter_Rotating_Review}. On the other hand,  BECs of attractively interacting atoms are only metastable under harmonic confinement, and they carry angular momentum via center-of-mass (COM) rotation rather than vortices \cite{2004_PRA_Baym_Attractive_Rotation}. 

Vortex lattices of two-component BEC mixtures were investigated for various intracomponent $g_{11}$, $g_{22}$, and intercomponent interactions $g_{12}$  \cite{2005_Ueda_Vortices, 2009_RMP_Fetter_Rotating_Review, 2001_PRL_Ho_Vortex, 2002_PRL_Mueller_Vortex_Jacobi_Theta}. Depending on the strength of the repulsive $g_{12}>0$, the mixture can exhibit coincident or displaced lattices with triangular, square, or rectangular symmetries. For the attractive intercomponent $g_{12}<0$ and mechanically stable $|g_{12}|< \sqrt{g_{11}g_{22}}$ Bose mixtures, both components coincide and form a triangular lattice \cite{2005_Ueda_Vortices}. However, the mean-field (MF) treatment of the Bose mixture problem predicts that the condensate is unstable towards collapse for $g_{12}<0$ and $|g_{12}|< \sqrt{g_{11}g_{22}}$ \cite{pethick2008bose}. The mixture can be metastable for a small number of particles if it is confined by a harmonic trap.

Quantum fluctuations can drastically modify the mean-field prediction toward collapse. Bose mixture quantum droplets are mechanically stable self-trapped phases without any external confinement, where attractive MF interaction $\propto -\delta g n^2$ is countered by the effectively repulsive beyond mean-field (BMF) fluctuations $\propto g_{BMF} n^{5/2}$ \cite{2015_Petrov_PRL}. Here,  $\delta g = |g_{12}| - \sqrt{g_{11}g_{22}}$ and $n$ is the condensate density.
For large particle numbers $N$, the interaction energy dominates the ground state, and the equilibrium density profile is almost like an incompressible liquid with a constant density \cite{2015_Petrov_PRL}.  For smaller $N$, the droplet can still stabilize itself, but the ground state exhibits a smoothly decaying density profile. There is a critical particle number below which the kinetic energy cost is too large and the droplet cannot sustain itself against expansion if it is not trapped.    

The quantum droplet has features of both the attractively interacting BEC due to mean-field and the repulsively interacting BEC due to fluctuations. The response of a quantum droplet to rotation is not straightforward. Specifically, the competition between the mean-field interaction and the quantum fluctuations gives rise to a new length scale, which is determined by the droplet's equilibrium density. Recent investigations of the problem mainly concentrated on weakly trapped condensates. Angular momentum is primarily carried by the COM rotation for the liquid-like flat-density droplets in weak external confinement \cite{2023_Arxiv_Kavoulakis_, 2020_JPB_Kavoulakis_Rotating_Droplet}. This way, the droplet conserves the ground-state interaction energy by preserving the flat-top density profile. By the same physical reasoning, the droplets experience splitting instability under a density perturbation, i.e., the droplet tends to break into smaller fragments \cite{2018_PRA_Malomed_Splitting,2018_PRA_Tarruell_Swirling_Superfluids,2019_PRL_Reimann_Rotating_Droplet,2020_PRE_Abdullaev_Instability_in_droplets}. 
However, vortices can be locally stable within the unconfined flat-top droplets if phase imprinted \cite{2018_PRA_Malomed_Splitting,2018_PRA_Tarruell_Swirling_Superfluids}  or the external confinement is adiabatically removed \cite{2019_PRL_Reimann_Rotating_Droplet}. Even weak confinement can help avoid splitting instability and make vortex states stable. A rotating droplet under harmonic confinement is recently predicted to show a combination of COM rotation and vortex states \cite{2020_JPB_Kavoulakis_Rotating_Droplet,2023_Arxiv_Kavoulakis_}.

The literature on quantum droplets mostly focuses on their self-trapping property, making the confinement potential mostly irrelevant. Here, we first classify the physical regimes for a trapped droplet by investigating the energy scales in the problem. 
Relative strengths of the potential and interaction energies and the interplay between the free equilibrium density and particle number decide the significance of the confining potential for the rotating problem. We argue that a strongly confined droplet can be driven into the rapidly rotating regime, forming a large vortex lattice. 

We study such rapidly rotating droplets in the strongly confined TF regime by numerical and approximate analytical methods. We numerically find that the confined droplet exhibits a triangular vortex lattice under rapid rotation. Strong confinement ensures that the splitting instability is avoided even when the rotation frequency is close to the trap frequency. The density profile of the vortex lattice closely follows the TF profile, with a convex peak at the center and a rapid fall at the TF radius. As the rotation frequency approaches trapping frequency $\tildeOmega \rightarrow 1$, the effective confinement $\propto (1-\tildeOmega^2)$ weakens, and the density profile of the vortex lattice converges to flat-top droplet density. Contrary to the diverging size of rapidly rotating repulsive BECs \cite{2004_PRL_Cornell_LLL_Experiment,2004_PRL_Dalibard_LLL}, the strongly confined TF droplet converges to a finite radius at the limit $\tildeOmega \rightarrow 1$. Thus, the physics of self-trapping plays an important role in the rapid rotation limit.

Our numerical simulations show that the vortex core sizes near the center and the edge of the droplet are noticeably different. We develop an approximate analytical formula for the density dependence of the core size. The vortex cores in the repulsive BECs scales $\zeta \propto 1/\sqrt{n_0}$, where $n_0$ is the condensate density. For the strongly confined TF droplet, however, we find the core size 
$\zeta \propto \frac{1}{\sqrt{n_0} \sqrt{n_0^{1/2} - n_c^{1/2}}}$, where the pole in the denominator $n_c$ is approximately equal to the equilibrium density of the unconfined droplet. The divergence at a finite density creates an observable difference in the core size in different regions of the vortex lattice. We also calculate the deviation of the vortex lattice from a perfect triangular lattice. While this deviation remains small, it is observable and agrees with numerical results. Finally, we argue that the rapidly rotating droplets present an opportunity to realize BECs where the number of particles is within an order of magnitude of the number of vortices in the condensate. While the GP equation and our assumptions about local energy of fluctuations are expected to break down in this limit, we argue that rapidly rotating droplets may facilitate reaching the strongly correlated regime.

This paper is organized as follows. In Section \ref{sec: Theory of Rotating Bose Mixture Droplet}, we 
introduce the effective GP equation of mixture droplets
and
tabulate different parameter regimes of the problem.  In Section \ref{sec: Strongly Confined Thomas-Fermi Regime}, we discuss the TF solution for the strongly confined droplet and compare it with the numerical results of the Gross-Pitaevskii (GP) equation. In Section \ref{sec: Core Size }, we obtain a formula for the vortex core size and compare it with the numerical solution. We also discuss the corrections to the uniform triangular lattice. In Section \ref{sec:Vortex Lattice with the Low Filling Factor}, we discuss the possibility of obtaining a vortex lattice with lower filling factors. In Section \ref{sec:Experimental Discussion and Conclusion}, we discuss the experimental parameters required to realize the suggested phenomena and outline future research directions. 

\section{\label{sec: Theory of Rotating Bose Mixture Droplet} Model and parameter regimes}

We consider a weakly interacting binary BEC with equal masses $m_1=m_2=m$ and wavefunctions $\Psi_1(\mathbf{r}) = \Psi_2(\mathbf{r})=\Psi(\mathbf{r})$, and equal number of atoms $N_1=N_2=N$. 
Intracomponent $s$-wave scattering lengths are also assumed equal and repulsive $a_{11}=a_{22}\equiv a>0$ but the intercomponent scattering length is negative $a_{12}<-a$. This gives a slightly attractive MF interaction such that effectively repulsive BMF energy becomes significant for the condensate's mechanical stability \cite{2015_Petrov_PRL}. 
%
%
We assume that the density gradient is sufficiently low throughout the condensate such that the local density approximation (LDA) holds \cite{2015_Petrov_PRL,2016_PRA_Santos_Dipolar_Droplet_Theory,2018_PRA_Ancilotto_LDA}. The droplet is confined radially with angular frequency $\omega_{\perp}$ and along the $z$-axis with an angular frequency $\omega_z$. The total energy functional of the mixture is: 
\begin{eqnarray}
E&=& \int dV  \Biggl\{  \frac{\hbar^2}{2 m}|\nabla \Psi|^2+\frac{1}{2} m w_z^2 z^2|\Psi|^2+\frac{1}{2}mw_{\perp}^2 r^2 |\Psi|^2 \nonumber \\
&+&\left.\frac{g(1-\alpha_s)}{4}|\Psi|^4+\frac{8 \sqrt{2}}{15 \sqrt{\pi}}   g a^{3/2}(1+\alpha_s)^{5 / 2} |\Psi|^5\right\},
\label{Energy_Functional_Confined_Droplet}
\end{eqnarray}
where $\mathbf{r} = (x,y)$ is the position vector, $\alpha_s = |a_{12}|/a$ is the ratio of the scattering lengths, and $g = 4 \pi \hbar^2 a /m$ is the coupling constant. The first and second terms in the second line of Eq.~\eqref{Energy_Functional_Confined_Droplet} are the MF and BMF energies, respectively. Note that the MF energy becomes attractive when $\alpha_s$ becomes slightly greater than one. In the experiments, $\alpha_s$ can be tuned via Feschbach resonance. 

We assume tight confinement in $z$-direction and integrate the energy functional by assuming a Gaussian wavefunction $\phi_0(z) = \frac{1}{\left(\pi a_z \right)^{1/4}} e^{-z^2 / 2 a_z^2}$, with $a_z = \sqrt{\frac{\hbar}{m \omega_z}}$, i.e $\Psi(\mathbf{r},z) = \psi(\mathbf{r}) \phi_0(z)$.  While tight confinement in the third direction may seem to contrast with the LDA assumption, recent works have shown that the transition from three- to two-dimensional behavior is smooth
\cite{2018_PRA_Petrov_3to2_BMF_Correction,2022_PRA_Gajda_Droplet_2d}.
The integrated form of the local energy functional in Eq.~\eqref{Energy_Functional_Confined_Droplet} is qualitatively similar to the strictly two-dimensional LDA functional where energy functional has the form $|\Psi|^4 \ln(|\Psi|^2/\sqrt{e})$  \cite{2019_PRL_Reimann_Rotating_Droplet,2018_PRA_Tarruell_Swirling_Superfluids}. 
They both feature a shallow minimum below zero and a rapid rise after becoming positive.  
We also expect that integrating out the $z$-axis provides better quantitative agreement with the typical experimental confinement scenarios. In a frame rotating with angular velocity $\mathbf{\Omega} = \Omega \hat{e}_z$, minimization of the functional $E-\Omega L_z$ \cite{1980_Landau_Liftshitz_Stat_Mech_2}, with $L_z$ being angular momentum,  
gives the dimensionless extended GP equation scaled with the radial trapping frequency $\omega_{\perp}$ as  
\begin{eqnarray}
\tilde{\mu} \tilde{\psi}&=&-\frac{1}{2}(\mathbf{\tilde{\nabla}}-\mathbf{\tilde{\Omega}} \times \mathbf{\tilde{r}})^2 \tilde{\psi}+\frac{1}{2} \tilde{r}^2\left(1-\tilde{\Omega}^2\right) \tilde{\psi} 
\label{GP_Equation_Rotating_Droplet}
\\
&+&\gamma_{int}\left[(1-\alpha_s)|\tilde{\psi}|^2+\beta_{LHY}|\tilde{\psi}|^3\right] \tilde{\psi}, \nonumber 
\end{eqnarray}
where $\mathbf{\tilde{r}} = \mathbf{r}/a_{\perp}$, $\mathbf{\tilde{\nabla}} = a_{\perp} \nabla$, $\tilde{\psi} = a_{\perp} \psi$, $\tilde{\Omega} = \Omega/\omega_{\perp}$. The interaction parameters are $\gamma_{int} \equiv  \sqrt{2 \pi} \frac{a}{a_z}$  and $\beta_{LHY} \equiv \frac{16 \sqrt{2}}{3 \sqrt{5} \pi^{3 / 4}}\frac{a^{3 / 2}}{a_z^{1 / 2} a_{\perp}} (1+\alpha_s)^{5 / 2}$, where $a_{\perp} = \sqrt{\frac{\hbar}{m \omega_{\perp}}}$. 
Note that in the rapid rotation limit $\tildeOmega \rightarrow 1$, the gas is still mechanically stable due to the balance between mean-field attraction and LHY quantum fluctuations, in stark contrast with regular BECs.

Physical scales in strictly two-dimensional limit are determined by the oscillator length $a_\perp$, the inter- and intra-component scattering lengths $a_{12}$, $a$,  and 
the total particle number $N$.
The oscillator length $a_\perp$, 
giving the typical 2D cloud size in the trap is used as the unit of length. 
The confinement along $z$ creates an additional length scale $a_z$, which will be used to tune effective two-dimensional interaction $\gamma_{int}\sim a/a_z$. The ratio of inter-to-intracomponent interaction $\alpha_s=|a_{12}|/a$ controls the attractive mean-field interaction, and droplet forms when it is larger than 1 yet the gas remains dilute. Last but not least, the dimensionless parameter $\beta_{LHY}$ determines the importance of the Lee-Huang-Yang BMF repulsion compared with the MF attraction. Although the parameters $\alpha_s$, $\beta_{LHY}$, and $\gamma_{int}$ seem to be inter-dependent at first sight, each is a unique combination of $s$-wave scattering, and the two oscillator lengths, and we choose to consider them independent parameters in the following for simplicity.

The limit of dominant interactions over both the kinetic and potential energies is a particularly important regime for droplet physics. In this case, the GP equation in \eqref{GP_Equation_Rotating_Droplet} reduces to: 
\begin{eqnarray}
\tilde{\mu} \tilde{\psi}=\gamma_{int}\left[(1-\alpha_s)|\tilde{\psi}|^2+\beta_{LHY}|\tilde{\psi}|^3\right] \tilde{\psi}.
\label{GP_Equation_TF}
\end{eqnarray}
The equilibrium density $n_{min}$ can be found by minimizing the term in the square brackets in \eqref{GP_Equation_TF} with respect to $|\tilde{\psi}|$ as  
\begin{eqnarray}
n_{min} = \frac{4(\alpha_s-1)^2}{9 \beta_{LHY}^2},
\label{n_min}
\end{eqnarray}
which is the density of a self-trapped droplet. 
In ``TF droplet'' regime, the kinetic energy is negligible and the density is flat if the trap potential is weak. Stronger trap potential squeeze the cloud to the center to minimize the confinement energy, which gives a convex peak at the center that rapidly falls near the surface. Outside the TF regime, which we call the ``weak droplet'' regime, the kinetic energy is non-negligible compared to the interaction energy, and the wavefunction gradient gradually smears the flat-top profile. Here, the droplet is still self-trapped but the density profile has deviations from $n_{min}$ due to the kinetic energy.

\begin{table}[b]
\caption{\label{tab:Different Regimes}%
Different regimes of the rotating confined droplet problem. The 
conditions
in Eqs.~\eqref{TF_regime} and \eqref{strong_confinement_regime} are used to determine the different regimes.   }
\begin{ruledtabular}
\renewcommand{\arraystretch}{1.5}
\begin{tabular}{ccc}
    & Strongly Confined & Weakly Confined \\
\hline
TF Droplet & 
$\frac{1}{\gamma_{int} |\alpha_s-1| N} \ll 1$ &
$\frac{1}{\gamma_{int} |\alpha_s-1| N} \ll 1$ \\
           &
$\frac{\beta_{LHY}^4}{\gamma_{int} |\alpha_s-1|^5} N \gtrsim 1$ &
$\frac{\beta_{LHY}^4}{\gamma_{int} |\alpha_s-1|^5} N \ll 1$      \\[5pt]
\hline
Weak Droplet &
$ \frac{1}{\gamma_{int} |\alpha_s-1| N} \approx 1$ & 
$ \frac{1}{\gamma_{int} |\alpha_s-1| N} \approx 1$ \\
             &
$\frac{\beta_{LHY}^4}{\gamma_{int} |\alpha_s-1|^5} N \gtrsim 1$ &
$\frac{\beta_{LHY}^4}{\gamma_{int} |\alpha_s-1|^5} N \ll 1$\\
\end{tabular}
\end{ruledtabular}
\end{table}

To quantify these regimes, we consider a constant density droplet with $\tilde{\psi}(\tilde{r}) = \sqrt{n_{min}} \mathcal \theta(R_{TF}-\tilde{r})$, where $\mathcal \theta$ is the Heaviside step function. For $N$ particles uniformly filling a circle of radius R, the scaling is obtained as 
$E_k \sim N / R^2$ for kinetic energy, 
$E_p \sim N R^2$ for potential energy, 
$E_{MF} \sim (N/R^2) N$ MF interaction, and 
$E_{LHY} \sim (N/R^2)^{3/2} N$ for fluctuations. 
However, due to the length scale corresponding to density $n_{min}$, the radius and the particle number are not independent. Expression $\int n_{min} d^2 \tilde{r} = N$ gives the TF radius of the flat-top droplet $R_{TF} = \frac{3 \beta_{LHY}}{2\sqrt{\pi}(\alpha_s-1)}\sqrt{N}$ and we find
\begin{eqnarray}
\label{energy_scales_Eint}
E_{int} &=& \frac{-14}{135} \frac{\gamma_{int} |\alpha_s-1|^3}{\beta_{LHY}^2} N, \\
\label{energy_scales_Ep}
E_p &=& \frac{9}{16 \pi} \frac{\beta_{LHY}^2}{|\alpha_s-1|^2} N^2, \\
E_k &=& \frac{2 \pi}{9} \frac{|\alpha_s-1|^2}{\beta_{LHY}^2}, 
\label{energy_scales_Ek}
\end{eqnarray}
where $E_{int}$ is the total interaction energy.
In the TF regime, the total interaction energy is much larger than the kinetic energy: 
\begin{eqnarray}
    \frac{E_k}{|E_{int}|} = \frac{30\pi}{14} \frac{1}{\gamma_{int} |\alpha_s-1| N} \ll 1,
    \label{TF_regime}
\end{eqnarray}
In the strong confinement regime, the potential energy is larger than the  interaction energy: 
\begin{eqnarray}
    \frac{E_p}{|E_{int}|} = \frac{1215}{224 \pi} \frac{\beta_{LHY}^4}{\gamma_{int} |\alpha_s-1|^5} N \gtrsim 1.
    \label{strong_confinement_regime}
\end{eqnarray}
Since $\gamma_{int}$ independently scales the interaction energy, one can drive the system into the weakly confined TF regime by increasing $\gamma_{int} \propto a/a_z$.  
Interaction energy \eqref{energy_scales_Eint} 
and 
the potential \eqref{energy_scales_Ep} 
scale with $N$ and $N^2$, respectively, whereas the kinetic energy \eqref{energy_scales_Ek} 
has a slower dependence (considering the full GP equation rather than the estimations Eq.~\eqref{energy_scales_Eint}-\eqref{energy_scales_Ek}). Therefore, the strongly confined TF regime can be achieved by large particle number $N$.
Note that
the condition \eqref{TF_regime} for the TF regime is independent of $\beta_{LHY}$, since both the interaction \eqref{energy_scales_Eint} and kinetic \eqref{energy_scales_Ek} energies scale with $\beta_{LHY}$. 
The different parameter regimes are summarized in Table \ref{tab:Different Regimes}.
The strongly confined TF regime,
where kinetic energy is negligible but potential energy causes a significant curvature in the equilibrium density, is eventually realized for all parameters if the number of particles is large enough.  

\section{\label{sec: Strongly Confined Thomas-Fermi Regime} Strongly Confined Thomas-Fermi Regime}

Rotating weakly confined droplet results in either COM motion that preserve the cloud density profile
\cite{2020_JPB_Kavoulakis_Rotating_Droplet,2023_Arxiv_Kavoulakis_},
or so-called splitting instability that divides the system into smaller fragments
\cite{2018_PRA_Malomed_Splitting,2018_PRA_Tarruell_Swirling_Superfluids}.
Stronger confinement prevents both of these undesirable effects and enables access to rapid rotation limits with stable vortex lattices. In this section, we focus on strongly confined rotating droplets in the TF regime and examine the coarse-grained density profile of the cloud as a function of the rotation frequency $\tilde{\Omega}$.

When a regular vortex lattice is formed, the kinetic energy of the rotating condensate is nearly equal to the energy of a solid body rotating with angular velocity 
$\mathbf{\Omega}$. 
The GP equation governing 
the coarse-grained density profile is found by neglecting the first term in \eqref{GP_Equation_Rotating_Droplet}: 
\begin{align}
\tilde{\mu} \tilde{\psi}&-\frac{1}{2} \tilde{r}^2\left(1-\tilde{\Omega}^2\right)\tilde{\psi} \nonumber\\
&= \gamma_{int}\left[(1-\alpha_s)|\tilde{\psi}|^2+\beta_{LHY}|\tilde{\psi}|^3\right] \tilde{\psi},
\label{GP_Equation_Rotating_Droplet_Strongly_Confined}
\end{align}
which manifests that the rotation affects the density profile mainly through centrifugal potential. The particles are squeezed toward the center for strong confinement,
whereas the centrifugal potential soften
the effective trapping frequency $(1-\tilde{\Omega}^2)$.
 
\begin{figure}[t]
    \includegraphics[width=0.23\textwidth,keepaspectratio]{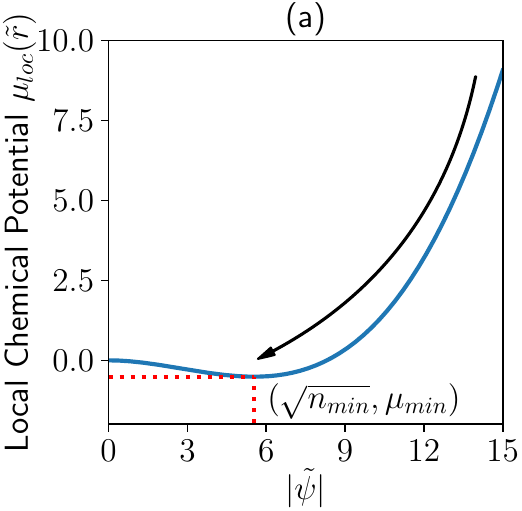}
    \includegraphics[width=0.23\textwidth,keepaspectratio]{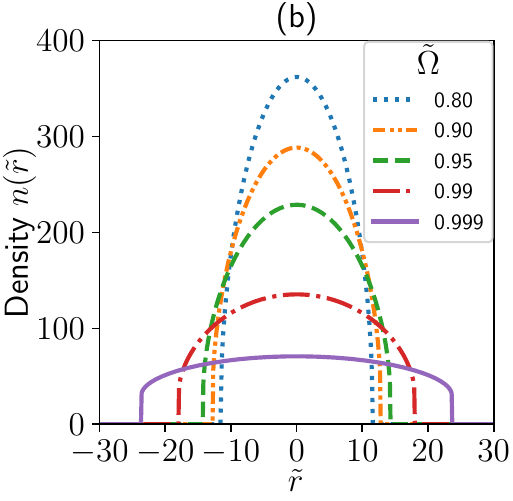}
    \caption{(a) The local chemical potential $\mu_{loc}(\tilde{r}) = \tilde{\mu}-\frac{1}{2} \tilde{r}^2\left(1-\tilde{\Omega}^2\right)$ in \eqref{GP_Equation_Rotating_Droplet_Strongly_Confined} vs. $|\tilde{\psi}|$ (Blue line). Since $\alpha_s >1$ and $\beta_{LHY}$ is positive, $\mu_{loc}$ shows a local negative minimum at $|\tilde{\psi}| = \sqrt{n_{min}}$, then monotonically increase with $|\tilde{\psi}|$. The solution to \eqref{GP_Equation_Rotating_Droplet_Strongly_Confined} for $\tilde{\mu}> \tilde{\mu}_{min}$ can be found by evaluating $\mu_{loc}(\tilde{r})$ for different $\tilde{r}$, along the curved arrow. The density jumps to 0 when the local chemical potential is equal to the minimum value $\mu_{min}$. (b) The analytical solutions of the GP equation \eqref{GP_Equation_Rotating_Droplet_Strongly_Confined} of the rotating Bose mixture droplets in the strongly confined TF regime for various $\tilde{\Omega}$. The parameters are $\alpha_s = 1.05$, $\beta_{LHY} = 6 \times 10^{-3}$, $\gamma_{int} = 1$, $N = 10^5$. The convexity due to the strong confinement flattens and approaches the flat-top profile as $\tilde{\Omega} \rightarrow 1$.} 
    \label{fig:RHS_vs_psi}
\end{figure}

As shown in Fig.~\ref{fig:RHS_vs_psi}a, the local chemical potential $\tilde\mu_{loc}$ obtained from \eqref{GP_Equation_Rotating_Droplet_Strongly_Confined} as a function of $|\tilde{\psi}|$ has negative minimum $\tilde{\mu}_{min}$ at $|\tilde{\psi}| = \sqrt{n_{min}}$, since $\alpha_s>1$ and $\beta_{LHY}>0$.
Now consider the solution $|\tilde\psi|$ of  \eqref{GP_Equation_Rotating_Droplet_Strongly_Confined} for a given chemical potential $\tilde{\mu}>\tilde{\mu}_{min}$.
At the center of the trap $\tilde{r} = 0$, $|\tilde{\psi}|$ takes its maximum value. As $\tilde{r}$ increases, the value of $|\tilde{\psi}|$ follows the path shown by the curved arrow in Fig.~\ref{fig:RHS_vs_psi}a. For the radial position $\tilde{r} = R_0$ at which $|\tilde{\psi}(R_0)| = \sqrt{n_{min}}$, the local chemical potential $\tilde{\mu}-\frac{1}{2} \tilde{r}^2\left(1-\tilde{\Omega}^2\right)$ reaches its minimum, and then $|\tilde{\psi}(\tilde{r})|$ abruptly falls to zero for $\tilde{r}>R_0$.  
We solve the cubic equation analytically as shown in App.~\ref{app:cubic_equation}. The solutions of Eq.~\eqref{GP_Equation_Rotating_Droplet_Strongly_Confined} for different rotation frequencies $\tildeOmega$ are shown in Fig.~\ref{fig:RHS_vs_psi}b, which demonstrate that the cloud flattens as the rotation frequency approaches the limit $\tilde{\Omega}\rightarrow 1$. 

We verify the above qualitative description by solving the nonlinear GP equation \eqref{GP_Equation_Rotating_Droplet} fully numerically in the strongly confined TF regime for the rapid rotations. 
We find the ground state in the rotating frame by the imaginary time evolution iterated by the split-step Fourier method \cite{2006_Bao_Split_Step,2006_Arxiv_Split_Step_Fourier}. We consider a typical choice of strongly confined TF regime parameters: $\alpha_s = 1.05$, $\beta_{LHY} = 6 \times 10^{-3}$, $\gamma_{int} = 1$, and $N = 10^5$. This choice gives  
$\frac{E_k}{|E_{int}|} = 1.37\times 10^{-4}$, 
$\frac{E_p}{|E_{int}|} = 2.3\times 10^{3}$,
$n_{min} \approx 30$ and TF radius $R_{TF} \approx 32$.  
\begin{figure*}[t]
    \includegraphics[width = 0.8\textwidth]{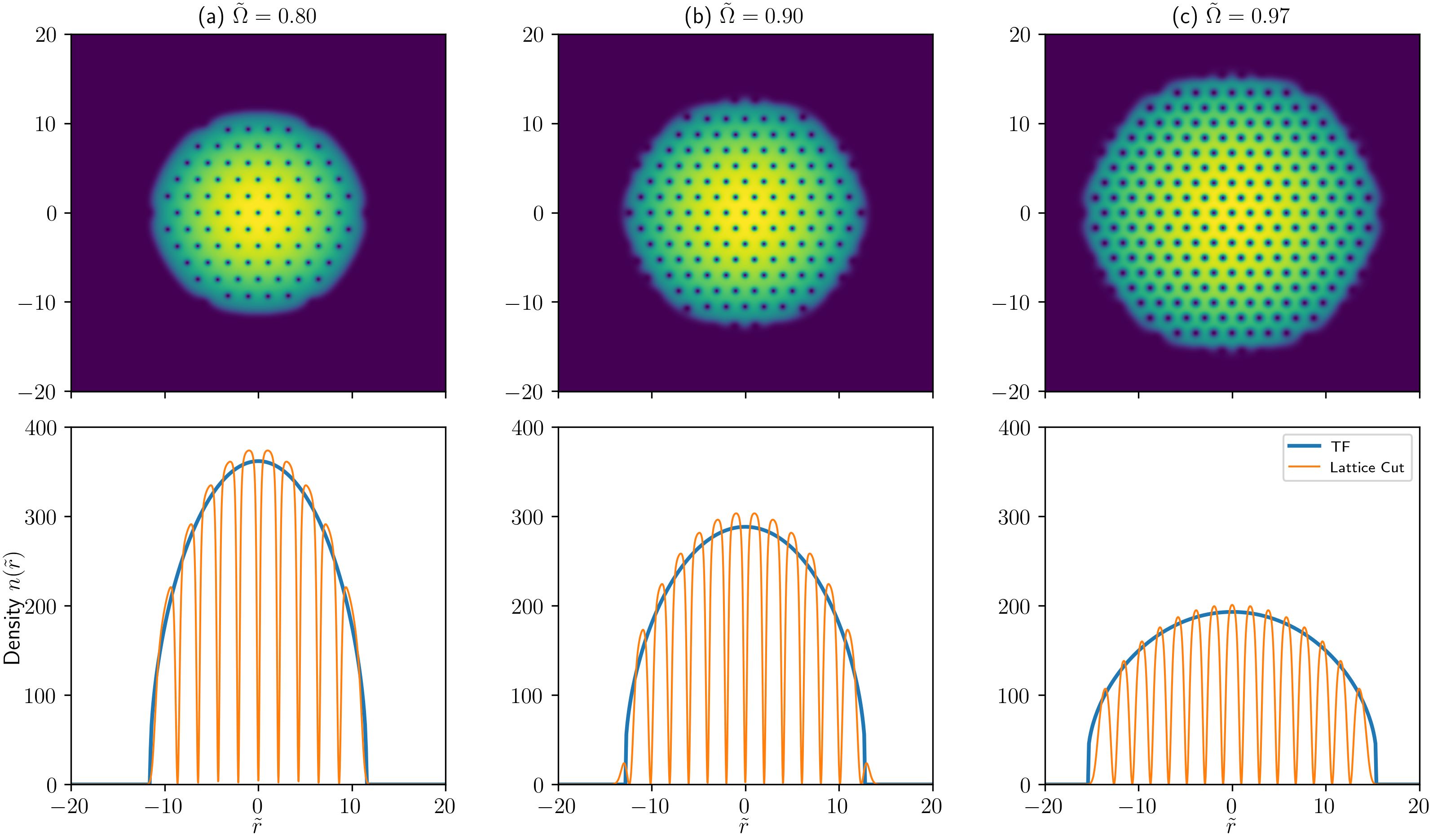}
    \caption{The numerical solutions of the GP equation \eqref{GP_Equation_Rotating_Droplet} of the rotating Bose mixture droplets for the parameters $\alpha_s = 1.05$, $\beta_{LHY} = 6 \times 10^{-3}$, $\gamma_{int} = 1$, $N = 10^5$, and (a) $\tilde{\Omega} = 0.80$,(b) $\tilde{\Omega} = 0.90$,(c) $\tilde{\Omega} = 0.97$. (upper panel) 2-D density profiles of the vortex lattices, (lower panel) the density profile of the vortex lattices along the $x$-axis (orange lines), and the rotating TF solution of the GP equation $\eqref{GP_Equation_Rotating_Droplet_Strongly_Confined}$ (blue lines).}
    \label{fig:Vortex_Lattices}
\end{figure*}
We tested the convergence of the numerical routine by using random, Gaussian, and Jacobi-theta initial wavefunctions \cite{2002_PRL_Mueller_Jacobi_Theta}. Each resulted in almost identical density profiles and locally triangular vortex lattices with occasional dislocation defects. The displayed results in Fig.~\ref{fig:Vortex_Lattices} are for Jacobi-theta function initial conditions, which have a perfectly periodic vortex lattice with a single vortex per hexagonal unit cell. This choice accelerates the convergence to minimum energy considerably and prevents unwanted dislocations in the lattice.
The lower panel of Fig.~\ref{fig:Vortex_Lattices} shows the density profiles of the cloud, which follows the solutions of the TF equation \eqref{GP_Equation_Rotating_Droplet_Strongly_Confined} (solid lines) with remarkable accuracy. With increasing rotation, the peak density at the center decreases, and the overall profile flattens, whereas the kinetic energy makes a marginal influence by smoothing the jump at the edge of the cloud.

\begin{figure}[H]
    \includegraphics[width=0.45\textwidth]{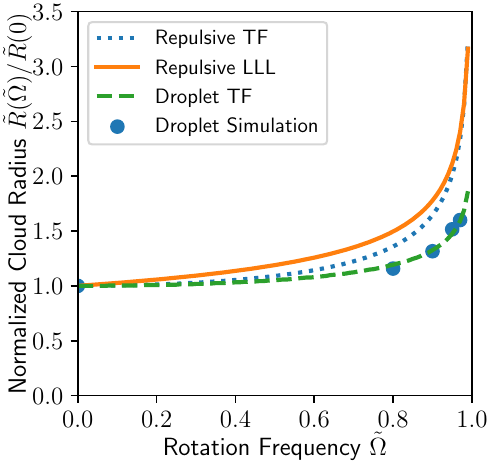}
    \caption{The comparison of the radius of the condensates for the rotation frequency $\tilde{\Omega} \in [0,1]$: Repulsive BEC in TF regime (dotted), Repulsive BEC in LLL regime (solid), and Droplet in the TF regime (dashed). The scattered data present the radius of the simulation results. The droplet radius converges to a finite value as $\tilde{\Omega} \rightarrow 1$, contrary to the diverging behavior of the repulsive BECs.} 
    \label{fig:radius_vs_omega}
\end{figure}

It is important to compare the course grained properties of the rapidly rotating droplet with the vortex lattices in the usual repulsive BEC experiments. 
In ordinary BECs, the TF profile always remains an inverted parabola, and its radius increases with increasing rotation. For droplets, there is a finite density jump at the surface and the functional form smoothly changes from a high curvature profile to a flat profile with increasing density.  
Notice that the cloud size of the rotating droplet for $\tilde{\Omega}=0.8$ and $\tilde{\Omega}=0.97$ are not substantially different in Fig.~\ref{fig:Vortex_Lattices}. In fact, as $\tilde{\Omega}$ increases towards one, the cloud radius converges to the non-rotating TF radius $R_{TF}$, whereas it diverges for the repulsive BECs in both TF and LLL regimes \cite{2004_PRL_Cornell_LLL_Experiment,2004_PRL_Dalibard_LLL,2002_PRL_Dalibard_Critical_Rotation,2001_PRL_Ho_Vortex}. This is a remnant of the celebrated self-trapping property of the droplets, where in this case, the fast rotation diminishes the external trapping potential but the competition between MF attraction and LHY repulsion constrains the increase of the cloud radius.
More concretely, for the repulsive BEC in TF and LLL regime, the cloud radius scales with $R(\tildeOmega) / R(0) = (1-\tildeOmega^2)^{-3/10}$ and $R(\tildeOmega) / R(0) = (1-\tildeOmega)^{-1/4}$, respectively \cite{2009_RMP_Fetter_Rotating_Review}. For the TF droplet, we calculate the radius for various $\tildeOmega$ from \eqref{GP_Equation_Rotating_Droplet_Strongly_Confined} in Appendix \ref{app:cubic_equation}, and also estimate it from the full numerical solutions of Eq.~\eqref{GP_Equation_Rotating_Droplet} as shown in Fig.~\ref{fig:radius_vs_omega}. One can see that the droplet radius remains much smaller than the regular BEC cloud radius, even in the extremely rapid rotation limit.

Most of the rotating BEC experiments were limited in the upper rotation frequencies due to the imperfections in the harmonic trap \cite{2008_cooper_rapidly_rotating_review}. These imperfections are harder to control away from the trap's center, and as the cloud radius gets larger, they dissipate angular momentum. We believe that the small change in the size of the droplet with rotation can make the rapid rotation limit easier to reach.

\section{\label {sec: Core Size } Properties of the Vortex Lattice }

The numerical results shown in Fig.~\ref{fig:Vortex_Lattices} reveal that the vortex core sizes are visibly different at the edge of the droplet compared with the center. 
This sharply contrasts with the repulsive BECs where the cores in the vortex lattices are almost uniform in size throughout the system \cite{1996_RPL_Ketterle}. In this section, we focus on the core sizes in the rapidly rotating TF droplet, and also the related lattice distortions previously known in the study of non-uniform superfluids, to better understand these observations.

\begin{figure}[t]
    \includegraphics[width=0.45\textwidth]{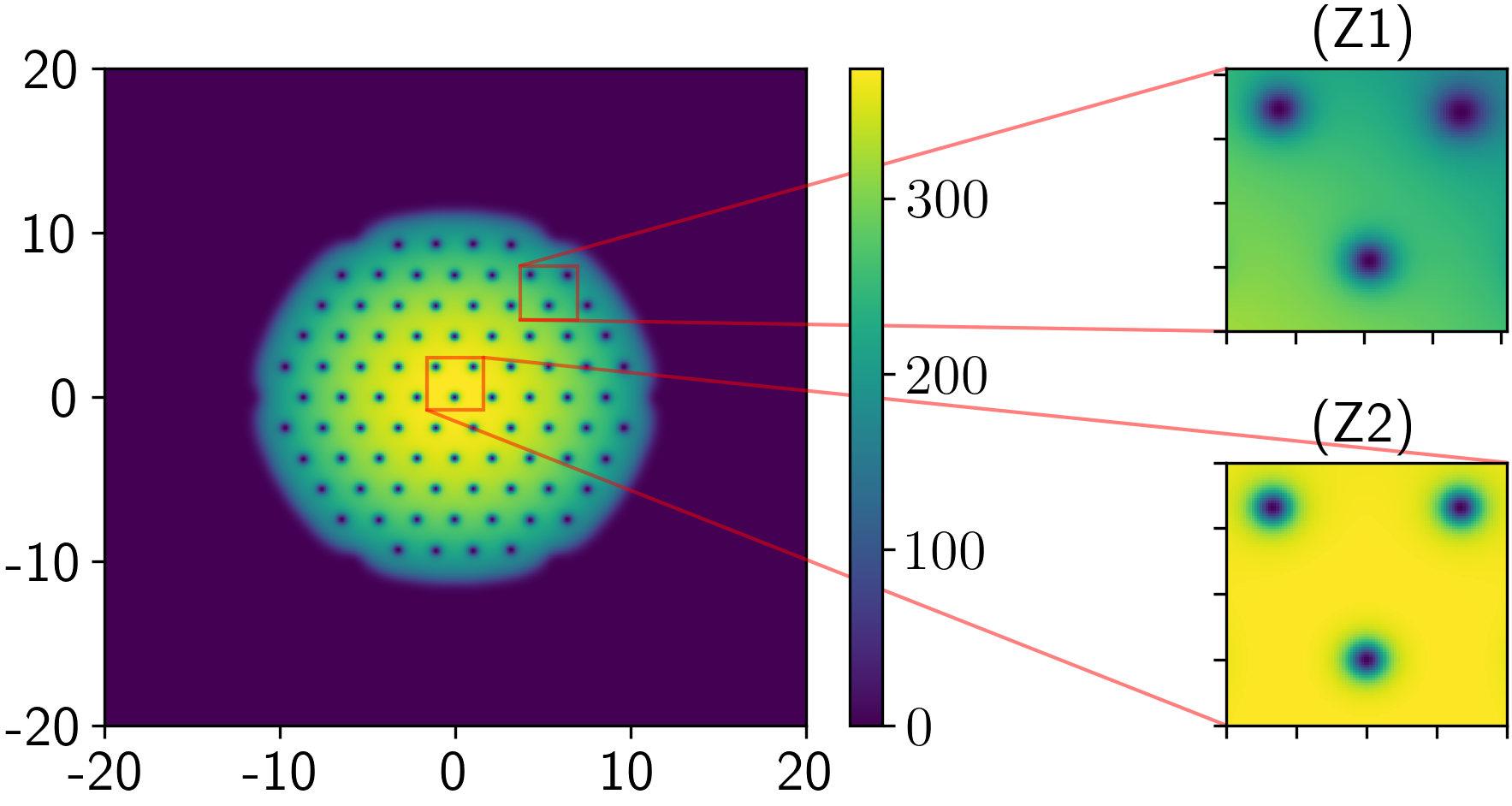}
    \caption{Density profile of the vortex lattice with $\tildeOmega = 0.8$ in Fig.\ref{fig:Vortex_Lattices}(a). The insets show the 2D zooms of equal areas in the vortex lattice near the edge (Z1) and the center (Z2) of the droplet. The core size is larger, and the vortex unit cell area decreases near the edge.} 
    \label{fig:lattice_zoom}
\end{figure}

Substituting the chemical potential $\tilde{\mu} = (1-\alpha_s)n_0 + \beta_{LHY}n_0^{3/2}$ corresponding to a uniform  
bulk
density $n_0$ in Eq.~\eqref{GP_Equation_TF}, one can obtain
\begin{align}
    \frac{1}{2} \tilde{\nabla}^2 \tilde{\psi}=
    (1&-\alpha_s)\left(|\tilde{\psi}|^2-n_0\right) \tilde{\psi}
    \nonumber\\
    &+\beta_{LHY}\left(|\tilde{\psi}|^3-n_0^{3 / 2}\right) \tilde{\psi}. 
    \label{coherence_length_formula}
\end{align}

We consider a semi-infinite condensate filling the right half-plane near an impenetrable surface to reduce this equation to a one-dimensional form. Multiplying both sides of Eq.~\eqref{coherence_length_formula} by $\partial \tilde{\psi}/\partial \tilde{x}$ and assuming a positive real wavefunction $\tilde{\psi}(\tilde{x}) \ge 0 $ for $\tilde{x}\ge 0$, we obtain

\begin{equation}
    \frac{1}{2}\frac{\partial }{\partial \tilde{x}} 
    \left(\frac{\partial \tilde{\psi}}{\partial \tilde{x}}\right)^2 - \frac{\partial V}{\partial \tilde{\psi}} \frac{\partial \tilde{\psi} }{\partial \tilde{x}} = 0,
\end{equation}
where
\begin{align}
    V[\tilde{\psi}] = (1&-\alpha_s)\left(\frac{1}{2}\tilde{\psi}^4 -n_0 \tilde{\psi}^2\right) \nonumber\\
    &+ \beta_{LHY} \left(\frac{2}{5}\tilde{\psi}^5-n_0^{3/2} \tilde{\psi}^2 \right).
\end{align}
This shows that $\frac{1}{2}(\partial \tilde{\psi}/\partial \tilde{x})^2- V[\tilde{\psi}]$ is a constant of motion. 
Evaluating this constant for the limits $\tilde{x} \rightarrow 0 $ and $\tilde{x} \rightarrow \infty$, where $\lim_{\tilde{x} \rightarrow 0,\infty}{\tilde{\psi}} = 0, \sqrt{n_0}$, gives the healing length scale of the condensate $\zeta \approx \sqrt{n_0}/\lim_{\tilde x\rightarrow 0}\left( \frac{\partial \psi }{\partial \tilde{x}}\right)$: 
\begin{eqnarray}
    \zeta &=&  \frac{1}{\sqrt{\frac{6\gamma_{int} \beta_{LHY}}{5} n_0 \left(n_0^{1/2}-n_c^{1/2}\right)}},
    \label{coherence_length_analytical}
\end{eqnarray}
where $n_c = \frac{25 (\alpha_s-1)^2}{36\beta^2_{LHY}}$ is the droplet density scale $n_{min}$ up to a numerical factor of order one. The analytical approximation fails for densities $n_0$ below $n_c$, signaling the splitting instability. For a better comparison, we also calculate the coherence length by numerically solving Eq.~\eqref{coherence_length_formula} in 1D and 2D geometries with appropriate boundary conditions, and extract independent estimations from the simulations based on
the imaginary time evolution iterated by the split-step Fourier method.
\begin{figure}[t]
    \includegraphics[width=0.22\textwidth,keepaspectratio]{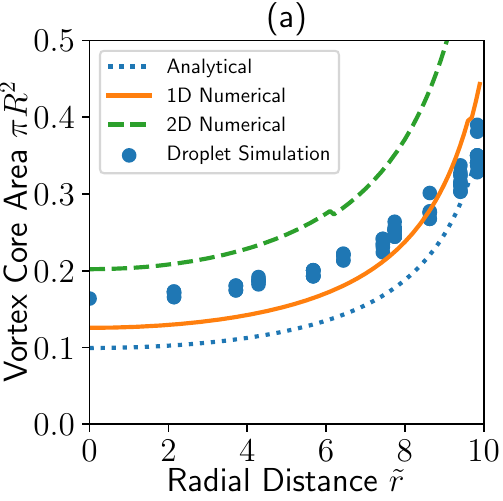}
    \includegraphics[width=0.22\textwidth,keepaspectratio]{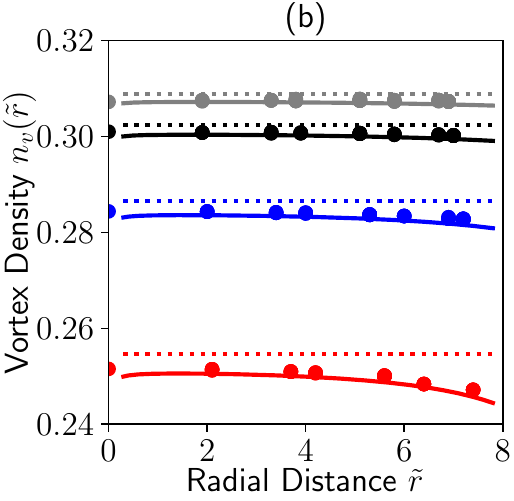}
    \caption{(a) The vortex core area estimates by numerical solutions of 1-D (solid line) and 2-D (dashed line) healing length equation \eqref{coherence_length_formula}, analytical approximation \eqref{coherence_length_analytical} (dotted line) and the simulation data for (filled points) $\tildeOmega = 0.8$. The density levels $n_0$ for the numerical and analytical solutions to \eqref{coherence_length_formula} and \eqref{coherence_length_analytical} are taken from the strongly confined TF droplet equation \eqref{GP_Equation_Rotating_Droplet_Strongly_Confined}. The difference in the core size along the TF droplet is more tractable than the repulsive BECs. (b) The vortex density of the rotating TF droplet calculated up to the first order in \eqref{Radziovsky_Vortex_Density} for $\tildeOmega = 0.80$ (gray), $\tildeOmega = 0.90$ (black), $\tildeOmega = 0.95$ (blue), and $\tildeOmega = 0.97$ (red). The dotted lines represent the uniform vortex densities $n_{v0} = \tildeOmega/\pi$. The scattered data shows the vortex density computed on the simulation data of Fig.~\ref{fig:Vortex_Lattices}, whereas the solid lines are the solutions to \eqref{Radziovsky_Vortex_Density}. The vortex density closely follows the uniform limit with an agreement on the corrections. } 
    \label{fig:core_size}
\end{figure}
As shown in Fig.~\ref{fig:core_size}, both the numerical estimates and the analytical result using the TF density expression for $n_0$ agree with the simulation data. The core size at the center of the droplet is smaller with respect to the surface of the cloud since the density of condensate decreases towards the edge.
%

In the rapidly rotating repulsive BECs in the TF regime, the core size $\zeta \propto \sqrt{\tildeOmega/2gn_0}$ \cite{2009_RMP_Fetter_Rotating_Review} and the TF density profile is inverse-parabola. For the TF droplet, the core size scales with the form given in Eq.~\eqref{coherence_length_analytical}, and the density is still finite $n_{min}$ near the surface. The density scale $n_c$ at the denominator gives a greater sensitivity of the core size to the density changes in the condensate. Furthermore, the finite density near the surface makes the vortices near the surface more visible compared to the surface of the inverse-parabola profile. 
 We expect that the differences in the vortex cores at the center and near the surface of the condensate should be more observable for the TF droplets with respect to the repulsive BECs.

Inhomogeneous superfluid density causes deviations in the local lattice constant, and the vortex density of an infinite uniform lattice given by Feynman relation $n_{v0} = \tildeOmega/\pi$ changes slightly. 
For rotating droplets, we numerically calculate this local change in the triangular lattice using image processing tools on simulation data and observe similar deviations. 
To estimate such changes in the vortex density, we follow the approach in Ref.~\cite{2004_PRA_Radziovsky_Vortex_Density}: 
\begin{eqnarray}
    n_v(\tilde{r})=\frac{\tilde{\Omega}}{\pi}+\frac{1}{8 \pi} \tilde{\nabla} \cdot \left\{\frac{1}{n(\tilde{r})}  \tilde{\nabla}\left[n(\tilde{r}) \ln \left(\frac{e}{\pi \zeta^2(\tilde{r}) n_v(\tilde{r})}\right)\right]\right\} \nonumber \\
    \label{Radziovsky_Vortex_Density}
\end{eqnarray}
where $n_v(\tildeOmega)$ is the vortex density, $n(\tilde{r})$ is the condensate density, and $\zeta(\tilde{r})$ is the coherence length. The first term in \eqref{Radziovsky_Vortex_Density} is the uniform vortex density $n_{v0}$ in dimensionless form, and the second term is the aforementioned small correction. 
We find the value of Eq.~\eqref{Radziovsky_Vortex_Density} numerically in the strongly confined TF droplet regime using the analytical superfluid density $n(\tilde{r})$ calculated in Appendix \eqref{Analytical_Solutiong_strongly_confined_GP}, and
the coherence length $\zeta(\tilde{r})$ calculated from \eqref{coherence_length_analytical} by replacing $n_0$ with  $n(\tilde{r})$. 
As the system is still well described by a single collective wavefunction, we expect the deviation between the superfluid density and the particle density to be small.
We compare the results with the vortex density profile of the simulation data for different values of $\tildeOmega$ in Fig.~\ref{fig:core_size}(b). 

The deviations from the uniform vortex density are due to the gradient of the superfluid density. Since the TF density profile becomes more flat as the rotation $\tildeOmega$ increases, the deviations from the uniform vortex density are greater for the slower rotations. Deviations become larger close to the edge of the droplet due to the convex profile. However, as $\tildeOmega \rightarrow 1$, the density profile approaches a flat-top shape, and the vortex density gets close to the uniform vortex density throughout the condensate. We expect an almost perfect triangular lattice for the highest rapid rotation rates, unlike the radially distorted vortex lattice in the repulsive TF BECs \cite{2004_PRA_Radziovsky_Vortex_Density}.

\section{\label{sec:Vortex Lattice with the Low Filling Factor} Vortex Lattices with Low Filling Factors}

Reaching the strongly correlated regime in rapidly rotating ultracold gases to mimic the physics of electrons in quantum Hall systems is one of the major challenges in atomic, molecular, and optical physics
\cite{2008_viefers_hall_physics_review,2008_cooper_rapidly_rotating_review}. 
Recently, the realization of fractional quantum Hall states are reported for photons and Bose gases but only with few particles \cite{2020_Nature_Clark_Photon_Laughlin_State,2023_Nature_Goldman_Fractiona_Quantum_Hall_State}. These experiments are far from simulating the full many-body physics at the mesoscopic scale, where some emergent properties, such as interaction-induced incompressibility, can be observed. 
Rapid rotation, which is the analog of the strong magnetic field in a neutral cold atomic gas,
effectively reduces the confining potential for 
the repulsive BECs in a harmonic trap. As $\tildeOmega$ increases, the condensate enters the MF LLL regime, where intervortex spacing becomes comparable to vortex core size. For a two-dimensional condensate uniform over a length $Z$ in the transverse direction, the cloud size is given by $R_0 = \left(\frac{8Na d_{\perp}^4}{Z (1-\tildeOmega)} \right)^{1/4}$ in the MF LLL regime \cite{2009_RMP_Fetter_Rotating_Review}, which diverges as $\tildeOmega \rightarrow 1$. 
As the inter-particle distance approaches the inter-vortex spacing, the quantum fluctuations become relevant, and the GP MF  description becomes insufficient
\cite{2008_cooper_rapidly_rotating_review}. One may still ask, however, whether some preliminary hints of many-body correlations emerge from the GP approach before it breaks down completely. 

The importance of correlations is characterized by the filling factor $\nu = N/N_v$, where $N$ and $N_v$ are particle and vortex numbers, respectively \cite{2000_PRL_Wilkin_Gunn_Composite_Bosons, 2001_PRL_Cooper_Phases_of_Vortices}. For $\nu \approx 5-10$, the quantum fluctuations are expected to melt vortex lattice and
drive the system into a fractional quantum Hall phase \cite{2002_PRL_MacDonald_Melting}. 
In the MF LLL regime, the number of vortices is given by
$
     N_v \approx R_0^2/a_{\perp}^2 
$
and the corresponding filling factor is \cite{2009_RMP_Fetter_Rotating_Review,2008_cooper_rapidly_rotating_review}:
 \begin{eqnarray}
     \nu \equiv \frac{N}{N_v}=\sqrt{\frac{Z(1-\tilde{\Omega}) N}{8 a}}.
 \end{eqnarray}
For typical values of $Z/a \approx 100 $, $N \approx 1000$, rotation rates $\tildeOmega > 0.99$ is needed to achieve $\nu \sim 5$. For such high values of $\tildeOmega$, the cloud size is much larger than the non-rotating condensate size. 

\begin{figure}[t]
    \includegraphics[width=0.4\textwidth]{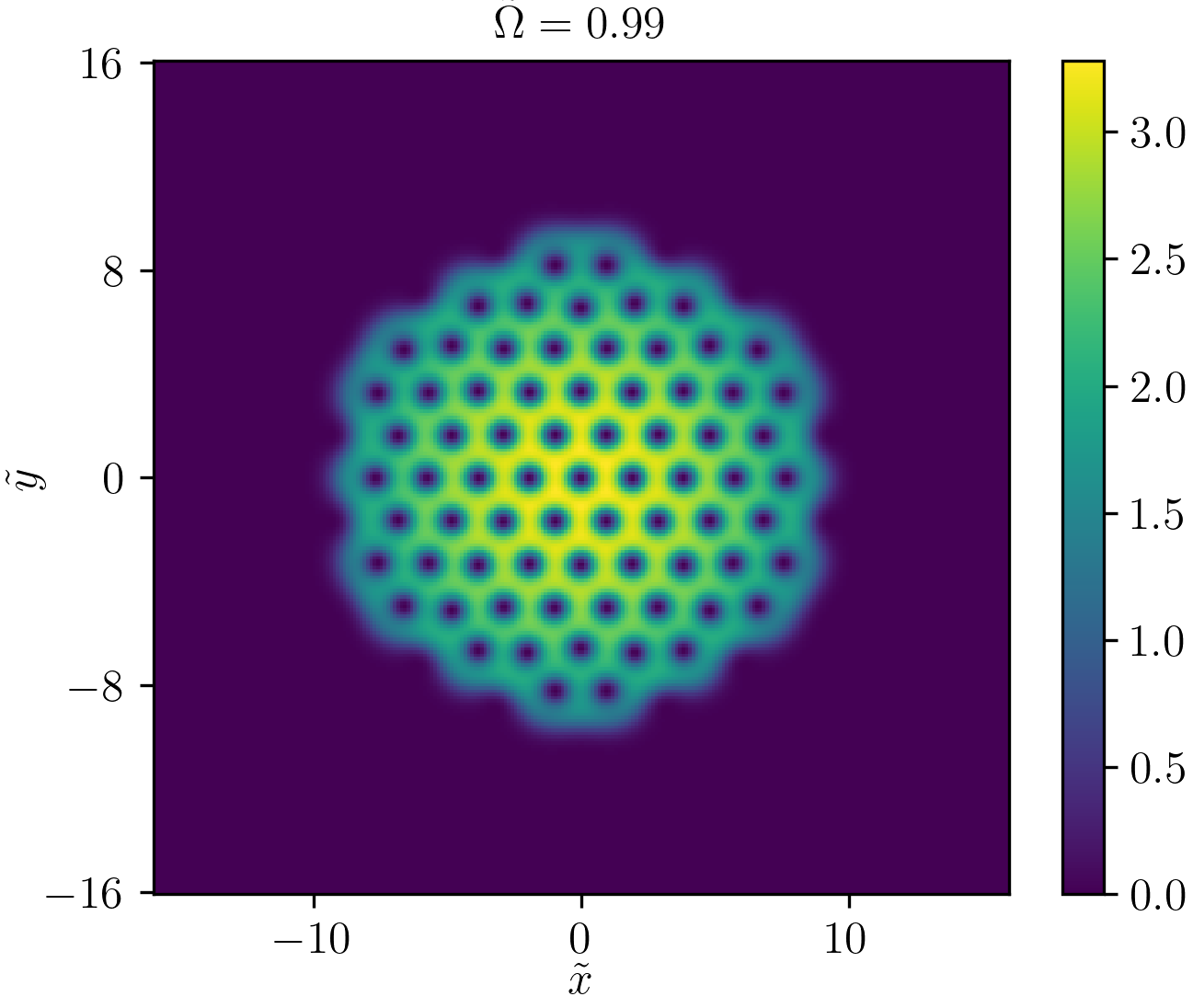}
    \caption{The vortex lattice of a rotating strongly confined droplet with a low filling factor. There are approximately 80 vortices for particle number $N= 500$, i.e. $\nu \approx 6.25$. The parameters are $\alpha_s = 1.25$, $\beta_{LHY} = 0.15$, $\gamma_{int} = 10$, and $\tildeOmega =0.99$. The colorbar indicates the condensate density.} 
    \label{fig:Filling_Factor}
\end{figure}

As discussed in the previous sections, a rotating droplet's size changes only fractionally in the strongly confined TF limit,
hinting at the possibility of achieving lower filling factors.    
Rotating droplet approaches a flat-top particle density $n_{min}$
whereas the approximate vortex density is $n_{v0} = \tildeOmega/\pi$. Assuming that the radius 
is close to the flat-top TF radius $R_{TF}$, 
the filling factor of the rotating droplet 
becomes:
\begin{eqnarray}
    \nu 
    = \frac{N}{\pi R_{TF}^2 n_v} 
    =\frac{\pi n_{min}}{\tildeOmega}.
\end{eqnarray}
This shows that lower filling factor can be achieved via smaller droplet density $n_{min}$. As an example, we
consider the parameters $N=500$, $\alpha_s = 1.25$, $\beta_{LHY} = 0.15$, $\gamma_{int} = 10$, which theoretically result in a strongly confined TF droplet with $n_{min} \approx 1.2$, $R_{TF} = 11.5$. For $\tildeOmega = 0.99$, the filling factor is close to $\nu \approx 3.8$. The full numerical solution of the GP equation \eqref{GP_Equation_Rotating_Droplet} is obtained for these parameters, and the 2D density profile of the vortex lattice is shown in  Fig.~\ref{fig:Filling_Factor},
which displays about $70$-$80$ vortices giving a filling factor $\nu \approx 6.25$. 
The difference between the numerical result and the theoretical expectation is due to the fact that the density profile is far from the flat-top regime for this parameter choice.

The vortex lattice in a droplet has two distinct instabilities in the rapid rotation limit. At low filling factors, quantum melting due to fluctuations destroys vortex lattice paving the way for the strongly correlated phases. The other possibility 
is that as the droplet density becomes flatter, a large portion of the density approach $n_{min}$ and vortex core sizes exceed the intervortex spacing.  This is the same mechanism of splitting as seen in weakly trapped droplets. 
As our above example shows, it is possible to get into the low filling factor regime while avoiding the weakly trapped regime.  We believe that the tunability of the equilibrium density, as well as the weak size change under rapid rotation, make droplets ideal systems to probe low filling factor physics.

\section{\label{sec:Experimental Discussion and Conclusion} Discussion, experimental realization and Conclusion}

The standard approach of using a modified GP equation to describe the system can break down in two ways. On one hand, condensate wavefunction $\Psi$ can be destroyed in a strongly correlated phase giving rise to a many-body phase without a local order parameter. On the other hand, the vortex lattice can strongly modify fluctuations so that the LDA assumption is violated even before the onset of the strongly correlated regime. We do not expect the GP equation to describe the system near the transition quantitatively. However, we believe that our calculations show that much lower filling factors than what has been observed are possible with droplets.

There are several intriguing questions that require further research. To achieve analog of electronic phases under ultra-strong magnetic fields, it becomes essential to import larger angular momenta into the system, resulting in lower filling factors. The central challenge in cold atom experiments, as discussed earlier, is the cancellation of trapping potential due to the centrifugal potential. 
Our work reveals that droplets can overcome this challenge through self-trapping, a consequence of the balance between attractive MF interactions and effectively repulsive LHY quantum fluctuations. 
It is important to note that a low filling factor does not necessarily imply strongly correlated physics for a droplet, as the GP formalism is still a mean-field approach.
Similarly, whether a rapidly rotating quantum droplet can be described within the LLL regime needs more investigation. The traditional condition for the LLL regime, $\mu \approx gn_0 \ll 2\hbar \omega_{\perp}$, is not applicable since the chemical potential of the self-trapped droplet is negative. 
Consequently, the stabilization of the cloud by quantum fluctuations and whether this stabilization mechanism still exists in the LLL regime should be explored.

The Bose mixture droplets without any trap are experimentally realized by several groups \cite{2018_Science_Tarruel_Mixture_Droplet,2018_RPL_Modugno_Mixture_Droplet,2019_PRR_Modugno_Droplet_Experiment,2021_PRL_Arlt_Droplet_Experiment}. Similar to our assumptions in this paper, the validity of the LDA treatment of the MF and LHY terms, and the physics of vortices and vortex lattices are theoretically established in the recent work on dipolar droplets \cite{2018_PRA_Macri_Rotating_Dipolar_Droplet,2019_PRA_Andrew_Rotating_Dipolar_Droplet,2020_PRA_Recati_Rotating_Dipolar_Droplet,2021_PRA_Reatto_Rotating_Dipolar_Droplet,2022_PRA_Santos_Rotating_Dipolar_Droplet,2022_PRL_Bisset_Rotating_Dipolar_Droplet}. 
Furthermore, the vortex lattices are experimentally obtained in rotating dipolar droplets \cite{2022_Nature_Ferlaino_Vortex_Lattices,2023_Arxiv_Ferlaiono_Vortices_in_dipolar_BEC}. It is promising for our proposal that the extended GP approach shows good agreement between the numerical estimates and the experimental results \cite{2022_Nature_Ferlaino_Vortex_Lattices,2023_Arxiv_Ferlaiono_Vortices_in_dipolar_BEC}. 

Let us consider the feasibility of an experiment to realize the vortex lattices in the Bose mixture droplets. 
The non-dimensional parameters which control the different regimes of the problem are $\alpha_s = \frac{|a_{12}|}{a}$, $\gamma_{int} \equiv  \sqrt{2 \pi} \frac{a}{a_z}$,  and $\beta_{LHY} \equiv \frac{16 \sqrt{2}}{3 \sqrt{5} \pi^{3 / 4}}\frac{a^{3 / 2}}{a_z^{1 / 2} a_{\perp}} (1 + \alpha_s)^{5/2}$. To obtain results close to the parameter choice of this paper, 
we consider a Bose mixture of $^{39}$K atoms in states $|1,0 \rangle $ and $|1,-1\rangle $ with total particle number $N= 10^5$. The scattering lengths are $a = 60 a_0$ and $a_{12} = -63a_0$, where $a_0$ is the Bohr radius. The radial and vertical trap frequencies are $\omega_{\perp}/2\pi = 10$ kHz and $\omega_{z}/2 \pi = 40$  kHz. The corresponding trap length scales are $a_z \approx 80$ nm and $a_{\perp} \approx 160$ nm. These conditions yield $1.5 - 2.5 $ $\mu $m size of the rotating droplet for $\tildeOmega \sim 0.80-0.97$.
Note that this particular choice of parameters is only illustrative. Similar phenomena can be observed in a  wide regime of parameters.  

In summary, we find that a triangular vortex lattice is obtained for a strongly confined TF droplet under rapid rotation. The overall density profile of the cloud follows the TF form even in the presence of the lattice. We investigated the lattice's vortex core size and vortex density and obtained good agreement between analytical and numerical results. The condensate size does not diverge at extreme rapid rotation $\tildeOmega \rightarrow 1$, due to self-trapping. This behavior can provide greater experimental feasibility to reach rapid rotation. Furthermore, the lattices with low filling factors can be achieved by tuning the droplet density. A more detailed investigation of the rapidly rotating droplet phase is required to determine if the fluctuation-induced stability mechanism extends to non-condensed bosonic phases such as the fractional quantum Hall states.

\acknowledgments
Recently, we became aware of a paper that studies the rotating Bose mixture droplets at the limit $\tildeOmega = 1$ \cite{2023_Arxiv_Qi_Rotating_Bose_Mixture_Droplet}. 

This work is supported by TUBITAK 2236 Co-funded
Brain Circulation Scheme 2 (CoCirculation2) Project No.
120C066 (A.K.). 

\appendix
\section{Analytical Solution of the Cubic TF Equation}
\label{app:cubic_equation}
 The cubic equation $ax^3 + bx^2 + cx + d = 0$ can be reduced into the depressed form $t^3 + pt + q = 0$ with the change of variables $t = x + \frac{b}{3a}$, where $p = \frac{3ac-b^2}{3a^2}$, and $q = \frac{2b^3-9abc+27a^2d}{27a^3}$. The real solution to the problem can be found in the trigonometric form. Now consider the GP equation \eqref{GP_Equation_Rotating_Droplet_Strongly_Confined} for the strongly confined TF droplet, which is a cubic equation with $c=0$. Apply the change of variables $t = |\tilde{\psi}| + \frac{1-\alpha_s}{\beta_{LHY}}$. The trigonometric solution becomes: 

 \begin{eqnarray}
    &&|\tilde{\psi}(\tilde{r} \le R_{TF})| = \frac{2(\alpha_s-1)}{3\beta_{LHY}} \\
    &&\times \left\{ \frac{1}{2} + \cos \left[ \frac{1}{3}\arccos \left(1 +\frac{27\beta_{LHY}^2 \mu_{loc}(\tilde{r})}{2\gamma_{int}(\alpha_s-1)^3}\right) \right] \right\} \nonumber
    \label{Analytical_Solutiong_strongly_confined_GP}
 \end{eqnarray}
 where the local chemical potential $\mu_{loc}(\tilde{r}) = \mu - \frac{\tilde{r}^2}{2}(1-\tildeOmega^2)$ and $R_{TF}$ is the TF radius of the strongly confined TF droplet, at where $|\tilde{\psi}(\tilde{R}_{TF})| = \sqrt{n_{min}}$. We calculate $R_{TF}$, where the density falls rapidly, by using $|\tilde{\psi}(\tilde{R}_{TF})| = \frac{2 (\alpha_s-1)}{3\beta_{LHY}}$:

 \begin{eqnarray}
     R_{TF} &=& \sqrt{\left( \mu + \frac{4\gamma_{int} (\alpha_s-1)^3}{27 \beta_{LHY}^2}\right) \frac{2}{1-\tildeOmega^2}} \nonumber \\
     &=& \sqrt{ \frac{2(\mu - \mu_{min})}{1-\tildeOmega^2}}
 \end{eqnarray}
 where $\mu_{min} = -\frac{4\gamma_{int} (\alpha_s-1)^3}{27 \beta_{LHY}^2}$ is the chemical potential of the flat-top droplet. Consider the limit $\tildeOmega \rightarrow 1$, in which the external trap is effectively removed by the centrifugal effect, and the droplet approaches the flat-top profile. Hence, the divergence of the droplet radius is prevented as $\lim_{\tildeOmega \rightarrow 1} \mu = \mu_{min}$. In fact, one should recover the flat-top radius $R_{TF} = \frac{3 \beta_{LHY}}{2\sqrt{\pi}(\alpha_s-1)}\sqrt{N}$ at the limit $\tildeOmega \rightarrow 1$.

\bibliography{refs}

\end{document}